\documentclass[pra,showpacs,onecolumn]{revtex4}
\usepackage{dcolumn}
\usepackage{bm}
\usepackage{amssymb}
\usepackage{amsmath}
\usepackage{amsfonts}

\usepackage[]{graphicx}
\begin{document}
\bibliographystyle{prsty}
\title{Reciprocity and optical chirality}
\author{ Aur\'elien Drezet $^{1}$, Cyriaque Genet $^{2}$}
\address{(1) Univ.~Grenoble Alpes, CNRS, Institut N\'{e}el, F-38000 Grenoble, France,(2) Laboratoire des Nanostructures, ISIS, Universit\'{e} de
Strasbourg, CNRS (UMR7006) 8 all\'{e}e Gaspard Monge, 67000
Strasbourg, France}
\begin{abstract}
This text was published as Chapter 2 in \textsl{Singular and Chiral Nano Plasmonics}, Edited by S.V. Boriskina, N. I. Zheludev, PanStanford Publishing, pp. 57-96 (2015).  
\end{abstract}

 \maketitle
\section{Introduction}
\indent Chirality (or handedness as the word stems from the greek
$\chi\epsilon\acute{\iota}\rho$ meaning hand), refers to the lack or
absence of mirror symmetry of many
systems~\cite{Landau,Fedorov,Hecht}. It is a fascinating property having important consequences
in every areas of science. For example it is connected to several
fundamental problems such as the apparition of life, the origin of
homochirality (that is of single handedness) of many
biomolecules~\cite{Masson}, and also to the asymmetry between left
and right handed fermions with respect to electroweak
interaction~\cite{Yang}. Historically, I.~Kant was one of the first
eminent scholar to point out the philosophical significance of
mirror operation. Already in 1783 in his celebrated ``Prolegomena to
any future metaphysics'' he wrote\\

 \emph{``Hence the difference between
similar and equal things, which are yet not congruent (for
instance, two symmetric helices), cannot be made intelligible by
any concept, but only by the relation to the right and the left
hands which immediately refers to intuition.''}\cite{Kant}.\\

W.~Thomson (Lord Kelvin), which was one of most important figure
of physics at the end of the XIX$^{\textrm{th}}$ century defined
more precisely chirality in the following way:\\

\emph{``I call any geometrical figure or group of points, chiral
and say it has chirality, if its image in a plane mirror, ideally
realized cannot be brought to coincide with it
self.''}\cite{Kelvin}\\

The interest of Kelvin for chirality is not surprising. He played
himself a critical role in the foundation of electromagnetism and
thermodynamics, and he was well aware of the work presented in 1896
by M.~Faraday on what is now known as the Faraday effect which is
intimately related to optical activity and chirality \cite{Faraday}.
In this context, it is interesting to observe that like pasteur
after him Faraday also had fruitless attempts to establish some
relations between electricity, chirality and light, it was a letter
from Thomson in 1845 that actually led Faraday to repeat his
experiments with a magnetic field and to discover non
reciprocal gyrotropy (i.e. magnetic optical rotation)!\\
Remarquably, this
description  of Kelvin provides an operational definition of chirality particularly suited to optics, as we illustrate below. \\
\indent In optics indeed, since the pioneer work of Arago
\cite{Arago} in 1811 and Biot in 1812 \cite{Biot}, chirality is
associated with optical activity (natural gyrotropy), which is the
rotation of the plane of polarization of light upon going through a
3D chiral medium such as a quartz crystal or an aqueous solution of
sugar. The first mathematical description of optical activity arisen
from the work of Fresnel in 1825~\cite{Fresnel} who interpreted
phenomenologically the effect in terms of circular birefringence,
that is as a difference in optical index for left and right handed
circularly polarized light (respectively written LCP and RCP)
passing through the medium. However, the intimate relationship
between optical activity and chirality became more evident after the
work of Pasteur~\cite{Pasteur,Flack} in 1848 concerning the change
in sign of the optical rotatory power for enantiomorphic solution of
left and right handed chiral molecules of tartaric acid. In 1874, Le
Bel and van't Hoff~\cite{Lebel,Hoff} related rotatory power to the
unsymmetrical arrangements of substituents at a saturated atom, thus
identifying the very foundation for stereochemistry. Since then,
optical activity, including circular birefringence and dichroism,
the so-called Cotton effect \cite{Cotton}, that is the difference in
absorbtion for LCP and RCP,
have become very powerful probes of structural chirality in a variety of media and environments.\\
\indent With the recent advent of metamarials, that are artificially
structured photonic media, a resurgence of
interest concerning optical activity is observed. Current inspirations can be traced back to the pioneer work of Bose~\cite{Bose} who, as early as
1898, reported on the observation of rotatory power for
electromagnetic microwaves propagating through a chiral artificial
medium (actually left and right handed twisted jute  elements). In
the context of metamaterials, Lindman in 1920~\cite{Lindman} (see
also Tinoco and Freeman in 1957~\cite{Tinoco}) reported a similar
rotatory dispersion effect through a system of copper helices in the giga-Hertz (GHz)
range. Very recently, and largely due to progress in micro and nano
fabrication technics, researchers have been able to taylor
compact and organized optically active metamerials in the GHz
and visible ranges. It was for instance shown, that planar chiral
structures made of \emph{gammadions}, i.e. equilateral crosses made of four bented arms, in
metal or dielectric, can generate optical
activity~\cite{Pendry,Papakostas,Schwanecke,Vallius,Gonokami,Canfield,Canfield2,Zhang,Decker,Plum,Rogacheva,add1,add2,add3,Gansel,Wegener,Zhao}
with giant gyrotropic factors~\cite{Gonokami,Decker,Plum,Rogacheva}.
Important applications in opto-electronics and also for refractive
devices with negative optical index for
RCP and LCP, have been suggested in this context~\cite{Pendry,Wegener,Zheludev}.\\
\indent These studies have raised an important debate on the genuine
meaning of planar chirality \cite{Papakostas,Schwanecke,Vallius}. Indeed, since intuitively a
two-dimensional (2D) chiral structure, which is by definition a system which can not
be put into congruence with itself until left from the plane, is
not expected to display any chiral optical characteristics due to
the fact that simply turning the object around leads to the
opposite handedness. More precisely, it was
shown that since optical activity is a reciprocal property (that
is obeying to the principle of reciprocity of Lorentz, see below), it
necessarily implies that reversing the light path through the
medium must recreate back the initial polarization state. However,
since the sense of twist of a 2D chiral structure changes when looking from
the second side, this polarization reversal is impossible. It would otherwise
lead to the paradoxical conclusion that a left and right
handed structure generates the same optical activity in
contradiction with the definition, finally meaning that optical activity must
vanish in strictly planar chiral systems. This behavior strongly contrasts with
what is actually observed for 3D chiral objects having a helicoidal
structure (like a quartz crystal \cite{Arago,Pasteur,Flack}, a
twisted jute element \cite{Bose}, or a metal helix
\cite{Lindman,Tinoco}), in full
agreement with the principle of reciprocity of Lorentz
\cite{Landau,Lorentz,Carminati,Barron1,Barron2} since the sense of twist of
an helix is clearly conserved when we reverse back the
illumination direction. The experimental observation of
optical activity in gammadion arrays forces one to conclude that such systems must present a form of hidden 3D
chirality which turns fully responsible for the presence of
optical rotation that rules over the dominant 2D
geometrical chiral character
possessed by the system.\\
Things could have stop here, but the understanding of 3D chirality
was recently challenged in a pioneering study where it was shown
that chirality has a distinct signature from optical activity when
electromagnetic waves interact with a genuine 2D chiral structure
and that the handedness can be recognized~\cite{Fedotov}. While the
experimental demonstration was achieved in the GHz (mm) range for
extended 2D structures (the so called fish-scale
structures\cite{Fedotov}), the question remained whether this could
be achieved in the optical range since the optical properties of
materials are not simply scalable when downsizing to the nanometer
level. Theoretical suggestions were provided to overcome this
difficulty by using localized plasmon modes excited at the level of
the nanostructures~\cite{Fedotov2}. Surface plasmons (SPs)
\cite{Barnes,Genet,Novotny} are indeed hybrid photon/electron
excitations which are naturally confined in the vicinity of a metal
structure. As evanescent waves, SPs are very sensitive to local
variations of the metal and dielectric environments~\cite{Barnes}.
This property was thus used to tune some 2D chiral metal structure
to optical waves. Two series of experiences made in the near
infrared~\cite{Fedotov3} with fishscale structure and in the visible
with Archimidian spirals~\cite{Drezet1,Gorodetskisub} confirmed the
peculiarity of genuine
2D chirality at the nanoscale. \\
\indent In this chapter, we will review some fundamental optical properties associated, in full generality, with chiral systems. An algebraic approach will allow us to reveal in a simple way the underlying connexions between the concepts of chirality and reciprocity from which global classes of chiral elements will be drawned. These classes will be described with the framework of Jones matrices, enabling a clear discussion on their respective optical properties. Finally, a few examples taken from our recent work will be discussed as illustrative examples of the relevant of planar chirality in the context of nanophotonics.

\section{The reciprocity theorem and the principle of path reversal}

\subsection{The Lorentz reciprocity relation}\label{Lorentz}
In order to understand the physical meaning and implications of the
different chiral matrices we will discuss we need to introduce the
principle of reciprocity of Lorentz~\cite{Lorentz,Carminati,Landau}.
First, we remind that, under the validity conditions of the paraxial
approximation, the properties of light going through an optical
medium are fully characterized by the knowledge of the $2\times 2$
Jones matrix
\begin{eqnarray} {J}:=\left(\begin{array}{cc}
J_{\textrm{xx}} & J_{\textrm{xy}}
\\ J_{\textrm{yx}}& J_{\textrm{yy}}
\end{array}\right),\label{jones}
\end{eqnarray} 
which ties the incident electric field
$\mathbf{E}^{\textrm{(in)}}=E^{\textrm{(in)}}_x \hat{\mathbf{x}}+
E^{\textrm{(in)}}_y \hat{\mathbf{y}}$ to the transmitted electric
field $\mathbf{E}^{\textrm{(out)}}=E^{\textrm{(out)}}_x
\hat{\mathbf{x}}+ E^{\textrm{(out)}}_y \hat{\mathbf{y}}$ (defined in
the cartesian basis $x,y$ of the transverse plane). This corresponds
to the transformation $\mathbf{E}^{\textrm{(out)}}=\hat{J}
\mathbf{E}^{\textrm{(in)}}$
where $\hat{J}$ is the operator associated with Eq. (\ref{jones}).\\
\indent From the point of view of \begin{figure}[h]
\centering\includegraphics[width=8cm]{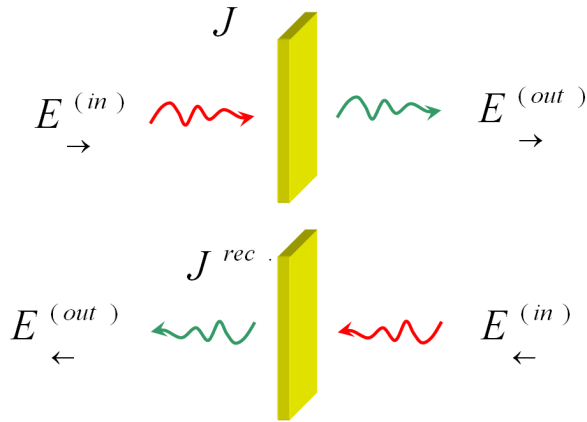} \caption{The two
different reciprocal histories for a light beam through a medium.
 The initial path going from left to right is represented by the Jones matrix $J$ while the reversed path which goes from right to left corresponds to the matrix $J^{rec.}$ (see Eqs. (\ref{eq5}) and (\ref{flip})). }\label{fig1}
\end{figure}
Jones matrices it is then possible
to give a simple formulation of the reciprocity principle. Consider
then a localized system illuminated by a plane wave
$\mathbf{E}^{\textrm{(in)}}_{\rightarrow}$ propagating along the
$+z$ direction (see Fig.\ref{fig1}). The arrow indicates the direction of
propagation. The transmitted field (in the paraxial approximation)
is given by
$\mathbf{E}^{\textrm{(out)}}_{\rightarrow}=\hat{J}\mathbf{E}^{\textrm{(in)}}_{\rightarrow}$.
Same, we define also an incoming plane wave propagating along the
$-z$ direction $\mathbf{E}^{\textrm{(in)}}_{\leftarrow}$ impinging
from the other side of the system and which after transmission gives
the output state
$\mathbf{E}^{\textrm{(out)}}_{\leftarrow}=\hat{J}^{\text{rec.}}\mathbf{E}^{\textrm{(in)}}_{\leftarrow}$.
Here by definition $\hat{J}^{\text{rec.}}$ is the reciprocal Jones
operator associated with $\hat{J}$. Using Maxwell equations one can
easily show that $\hat{J}^{\text{rec.}}=\hat{J}^{\text{T}}$ where
$T$ denotes the transposition in the cartesian basis
$\hat{\mathbf{x}},\hat{\mathbf{y}}$. The proof reads as follow:
first, from Maxwell equations one deduces \cite{Landau} the reciprocity
theorem of Lorentz which states that if in a passive and linear
environment we consider two space points $A$ and $B$ then the vector
field $\mathbf{E}(B)$ in $B$ produced by an (harmonic) point like
dipole source $\mathbf{P}(A)$ located in $A$ is linked to the field
$\mathbf{E}'(A)$ in $A$ produced by a second point like dipole
$\mathbf{P}'(B)$ located in $B$ through the formula
\begin{eqnarray}
\mathbf{P}(A)\cdot\mathbf{E}'(A)=\mathbf{P}'(B)\cdot\mathbf{E}(B) \label{eq2}
\end{eqnarray} 
(the time harmonic dependency $e^{-i\omega t}$ has been dropped
everywhere).\\
\indent Now, the electric field produced in $M$ by a point like
dipole located in $M'$ is written
$\mathbf{E}(M)=\underline{\mathbf{G}}(M,M')\cdot\mathbf{P}(M')$
where $\underline{\mathbf{G}}(M,M')$ denotes the dyadic Green
function for this environment~\cite{Novotny}. Equation (\ref{eq2}) reads
thus
\begin{eqnarray}
\mathbf{P}(A)\cdot(\underline{\mathbf{G}}(A,B)\cdot\mathbf{P}'(B))=\mathbf{P}'(B)\cdot(\underline{\mathbf{G}}(B,A)\cdot\mathbf{P}(A))\nonumber\\
\end{eqnarray} or equivalently in tensorial notation
$\sum\sum_{i,j}(G_{ij}(A,B)-G_{ji}(B,A))P_{i}(A)P_j'(B)=0$
($i,j=1,2,3$). This relation is valid for every point dipoles in $A$
and $B$ and implies consequently
\begin{eqnarray}
G_{ij}(A,B)=G_{ji}(B,A).
\end{eqnarray}
Actually, this relation constitutes a Maxwellian  formulation  of
the principle of light path reversal used in optical geometry. In
the next step of the proof we consider $A$ and $B$ located in $z=\pm
\infty$ the fields can be then considered asymptotically as plane
waves and in the paraxial approximation $G_{ij}(-\infty,+\infty)$
identifies with the Jones matrix $J_{ij}$. We immediately see that
the matrix $G_{ij}(+\infty,-\infty)$ identifies with the reciprocal
matrix $J^{\textrm{rec.}}_{ij}$. In other words, from the point of
view of Jones formalism, the principle of reciprocity states
\begin{eqnarray}
{J}^{\text{rec.}}:={J}^{\text{T}}=\left(\begin{array}{cc}
J_{\textrm{xx}} & J_{\textrm{yx}}
\\ J_{\textrm{xy}}& J_{\textrm{yy}}
\end{array}\right).\label{eq5}
\end{eqnarray}
\indent In this context it is relevant to point out the similarity
between the reasoning given here for establishing the reciprocity
theorem and the one used in textbooks and articles
\cite{Landau,Fedorov,Carminati} for establishing the symmetry of the
permittivity tensor $\epsilon_{i,j}$ ($i,j=1,2,3$) in solids. In
particular, by taking into account spatial non-locality it is
possible to obtain a version of the reciprocity theorem which reads:
$$\epsilon_{i,j}(\omega,-\mathbf{k})=\epsilon_{j,i}(\omega,\mathbf{k})$$ where
$\mathbf{k}$ is the wavevector of the monochromatic plane wave. The
analogy with Eq. (\ref{eq5}) is complete if we choose the wave vector along
the $z$ axis and if $i,j$ correspond to either $x$ or $y$. Because
of these similarities many reasoning done for the Jones matrix
through this chapter could be easily restated for the electric
permittivity $\epsilon$ or magnetic permeability $\mu$ tensors.
\subsection{Rotation of the optical medium
and reciprocity: conserving the handedness of the reference frame}
\indent Using the previous formalism the reciprocity principle give
us a univocal way to calculate the transmitted light beam
propagating in the $-z$ direction through a structure if we know the
transmission Jones matrix for propagation in the $+z$ direction.
However, we must remark that this formulation is not always the most
convenient since we compare the Jones matrix from a situation in
which the triplet of unit vectors built by
$\hat{\mathbf{x}},\hat{\mathbf{y}}$, and the wavevector $\mathbf{k}$
of the light wave  (along z in the paraxial regime) constitute a
right handed trihedra to a situation in which the same trihedra of
vectors is left handed (since the sign of $\mathbf{k}$ is opposite
in the two situation). This in particular means that for an observer
watching from one side $B$ of the medium a light beam coming from
the other side $A$ the definition for LCP and RCP light as
$\hat{\mathbf{L}},\hat{\mathbf{R}}=(\hat{\mathbf{x}}\pm i
\hat{\mathbf{y}})/\sqrt{2}$ is different from the one obtained by an
reciprocal observer watching from the side $B$ a light incident from
the $A$ side, i.e.,
$\hat{\mathbf{R}},\hat{\mathbf{L}}=(\hat{\mathbf{x}}\pm i
\hat{\mathbf{y}})/\sqrt{2}$. In order to remove this ambiguity the
principle of reciprocity can be reformulated  by considering  a
reference frame transformation which is a global flip of the optical
medium. More precisely, comparing directly ${J}^{\text{T}}$ to ${J}$
necessitates a rotation ${R}_x$ of the plane $y-z$ by an angle $\pi$
around $x$. Mathematically this three dimensional transformation
reads \begin{eqnarray}{R}_x=\left(\begin{array}{ccc} 1 & 0& 0\\ 0&
-1& 0\\0 &0& -1\end{array}\right).\end{eqnarray} The handedness of
the 3-axis coordinate system is kept unchanged after the application
of the reference frame transformation ${R}_x$. It also implies that
the full structure of Maxwell equation is also conserved, i.e., the
optical effect is rigorously equivalent to a $\pi$-rotation of the
system around $x$. In particular the wavevector
\begin{eqnarray}\mathbf{k}=\left(\begin{array}{c} 0\\0\\-|k|\end{array}\right)\end{eqnarray} of
the light going through the system transforms through ${R}_x$ into
\begin{eqnarray}\mathbf{k}'=\left(\begin{array}{c} 0\\0\\+|k|
\end{array}\right).\end{eqnarray}
\indent  From the point of view of the $2\times2$
Jones matrix ${R}_x$ simply reduces to ${\Pi_x}=\left(\begin{array}{cc} 1 & 0\\
0& -1\end{array}\right)$ that is to a planar mirror symmetry with
reflection axis parallel to $x$. Through ${R}_x$, the field vectors transform as $\mathbf{E}_{\Pi_x}=\hat{\Pi}_{x}\mathbf{E}$, so that the Jones matrix
becomes ${\Pi_x}\cdot{J}^{\textrm{T}}\cdot{\Pi_x}^{-1}$.\\
\indent Fundamentally it is interesting to note that a symmetry
operation, which is defined geometrically (and independently from
any set of physical laws), manifests itself specifically according
to a particular physical environment or context. From the point of
view of Maxwell's equations this symmetry operation is indeed
implemented at the level of the susceptibility of the medium
interaction with light that is at the level of the Jones matrices.\\
\indent Actually, the reciprocal transformation defined above constitutes a
rigorous and operational optical definition of the medium flipping
\begin{eqnarray}
{J}^{\textrm{flip.}}:={\Pi_x}\cdot J^{\textrm{rec}}\cdot{\Pi_x}^{-1}  \  \   {\textrm{with} }  \ Ê \ ÊJ^{\textrm{rec}} = {J}^{\textrm{T}}.\label{flip}
\end{eqnarray}
\indent It will be also convenient in the following to express the electric
field in the left (L) and right (R) circularly polarized light basis
defined by $\hat{\mathbf{L}},\hat{\mathbf{R}}=(\hat{\mathbf{x}}\pm i
\hat{\mathbf{y}})/\sqrt{2}$. We write $\overline{{J}}$ the Jones
matrix in such basis, and we have
\begin{eqnarray} \overline{{J}}=\left(\begin{array}{cc}
J_{ll} & J_{lr}
\\ J_{rl}& J_{rr}
\end{array}\right)= {U}{J}{U}^{-1},
\end{eqnarray}
where ${U}=\frac{1}{\sqrt{2}}\left(\begin{array}{cc} 1 & -i
\\ 1& i
\end{array}\right)$ defines the unitary matrix associated with this vector basis
transformation.\\
\indent With these definition Eq.~(\ref{flip}) reads in the RCP and
LCP basis
\begin{eqnarray}
 \overline{{J}}^{\textrm{flip.}}=\overline{{J}}^{\textrm{T}}=\left(\begin{array}{cc}
J_{ll} & J_{rl}
\\ J_{lr}& J_{rr}
\end{array}\right)  \label{flip2}
\end{eqnarray}
(remark that $\overline{{J}^{\textrm{T}}}=\left(\begin{array}{cc}
J_{rr} & J_{lr}
\\ J_{rl}& J_{ll}
\end{array}\right)\neq\overline{{J}}^{\textrm{T}}$
since the transposition $T$ and the transformation $\hat{U}$ are
not commutative operations).\\
\indent To summarize, we showed that ${{J}}^{\textrm{flip.}}$ and
$\overline{{J}}^{\textrm{flip.}}$ define the genuine representation
of reciprocity and path reversal in a coordinate system having  the
same handedness as the original one (i.e. as seen from the other
side of the object). This corresponds to the axes transformation
$x'=x$, $y'=-y$, and $z'=-z$ and it implies an exchange in the role
of LCP and RCP.\\
\subsection{Time-reversal versus reciprocity}
\indent We point out that the reciprocity relations (\ref{flip}) and (\ref{flip2}) should not be
confused with the time-reversal transform. Time-reversal is a
fundamental symmetry which dictates the invariance of physical laws
between exchange of past and future. In classical mechanics, this symmetry corresponds to a system described by its position and momentum $({\bf q},{\bf p})$ which equations of motion are invariant through the transformation $({\bf q},{\bf p},t)\rightarrow ({\bf q},-{\bf p},-t)$. While an isolated mechanical system is time-reversal, a real system is obviously always coupled with its environment, resulting in friction which immediately breaks the time-reversal symmetry.  \\
\indent In the presence of electromagnetic fields $\mathbf{E}(\mathbf{x},t)$,
$\mathbf{B}(\mathbf{x},t)$, this time-reversal invariance is
preserved if in addition to the variables $({\bf q},{\bf p})$ one also
transforms the field into
$\mathbf{E}'(\mathbf{x},t)=\mathbf{E}(\mathbf{x},-t)$,
$\mathbf{B}'(\mathbf{x},t)=-\mathbf{B}'(\mathbf{x},-t)$. In order
that Maxwell's equation be automatically satisfied by the  new
solutions $\mathbf{E}',\mathbf{B}'$ the electric current and charge
distributions are changed accordingly into
$\mathbf{J}'(\mathbf{x},t)=-\mathbf{J}(\mathbf{x},-t)$,
$\rho'(\mathbf{x},t)=\rho(\mathbf{x},-t)$. In the monochromatic
regime where fields, electric currents and charges are conveniently
described by their time-Fourier transforms at the positive pulsation
$\omega$ the time-reversal operation reads:
\begin{eqnarray}
\begin{array}{cc}
\mathbf{E}_\omega'(\mathbf{x})=\mathbf{E}^\ast_\omega(\mathbf{x}),&\mathbf{B}_\omega'(\mathbf{x})=-\mathbf{B}^\ast_\omega(\mathbf{x})
\\ \mathbf{J}_\omega'(\mathbf{x})=-\mathbf{J}^\ast_\omega(\mathbf{x}),&
\rho_\omega'(\mathbf{x})=\rho^\ast_\omega(\mathbf{x}).
\end{array}
\end{eqnarray}
The time dependence is restored after integration over the pulsation
spectrum e.g.,
$\mathbf{E}(\mathbf{x},t)=\int_0^{+\infty}{\rm d}\omega\mathbf{E}_\omega(\mathbf{x})e^{-i
\omega
t}+\int_0^{+\infty}{\rm d}\omega\mathbf{E}^\ast_\omega(\mathbf{x})e^{+i\omega
t}$ etc...\\
\indent For optical situations where a modal expansion into
plane-waves of vector
$\mathbf{k}=k_x\hat{\mathbf{x}}+k_y\hat{\mathbf{y}}$,
$k_z=\sqrt{(\frac{\omega}{c})^2\epsilon_\omega-\mathbf{k}^2}$ is
considered ($\epsilon_\omega$ being the complex-valued dielectric
permittivity of the medium) time-reversal implies new modal
components such as
\begin{eqnarray}
\begin{array}{cccc}
\epsilon'_\omega=\epsilon_\omega^\ast,&k'_z=k_z^\ast,&
\mathbf{E}'_{\omega,\pm}(\mathbf{k})=\mathbf{E}^\ast_{\omega,\mp}(-\mathbf{k}),&\mathbf{B}'_{\omega,\pm}(\mathbf{k})=-\mathbf{B}^\ast_{\omega,\mp}(-\mathbf{k}),
\end{array}
\end{eqnarray}
where by definition $\mathbf{E}_\omega(\mathbf{x})=\int
{\rm d}^2\mathbf{k}e^{i\mathbf{k}\mathbf{x}}
[\mathbf{E}_{\omega,+}(\mathbf{k})e^{ik_zz}+\mathbf{E}_{\omega,-}(\mathbf{k})e^{-ik_zz}]$
etc... In particular, if the medium is lossless, i.e.,
$\textrm{Imag}[\epsilon_\omega]=0$, the time-reversal operation
dictates in the propagative sector (i.e. for $|\mathbf{k}|\leq\omega
\sqrt{\epsilon_\omega}/c$) a change in the sign of the wave vectors
corresponding to a reversal of propagation direction for every
plane-waves of the modal expansion. Going back to the Jones matrix
formalism we can transform the relation
$\mathbf{E}^{\textrm{(out)}}_{\rightarrow}=\hat{J}\mathbf{E}^{\textrm{(in)}}_{\rightarrow}$
into
$\mathbf{E}^{\textrm{(in)},\ast}_{\rightarrow}=\hat{J}^{-1,\ast}\mathbf{E}^{\textrm{(out)},\ast}_{\rightarrow}$.  \\
\indent Now, from the previous discussion concerning time reversibility, the
complex conjugated input field
$\mathbf{E}^{\textrm{(in)},\ast}_{\rightarrow}$ at $z=-\infty$
corresponds to the time-reversed output
$\mathbf{E'}^{\textrm{(out)}}_{\leftarrow}$ computed at $z=-\infty$
whereas the complex conjugated output field
$\mathbf{E}^{\textrm{(out)},\ast}_{\rightarrow}$ at $z=+\infty$
corresponds to the time reversed output
$\mathbf{E'}^{\textrm{(in)}}_{\leftarrow}$ computed at $z=+\infty$.
The Jones matrix associated with time-reversal is therefore
\begin{eqnarray}
{J}^{\text{inv.}}:={J}^{-1,\ast}=\frac{1}{J^{\ast}_{\textrm{xx}}J^{\ast}_{\textrm{yy}}-J^{\ast}_{\textrm{xy}}J^{\ast}_{\textrm{yx}}}\left(\begin{array}{cc}
J^{\ast}_{\textrm{yy}} & -J^{\ast}_{\textrm{xy}}
\\ -J^{\ast}_{\textrm{yx}}& J^{\ast}_{\textrm{xx}}
\end{array}\right). \label{eq14}
\end{eqnarray}
\indent As it is clear from its definition, ${J}^{\text{inv.}}$ is in general
different from ${J}^{\text{rec.}}$, exemplifying the importance of losses and dissipation in the relation
between time reversibility and reciprocity in optics. The two operators are indeed
identical if, and only if, $J$ is unitary, i.e., ${J}^{-1}=J^\dagger$, meaning that an optical system through which energy is conserved and which is simultaneously reciprocal will be the only optical system to be time-reversal invariant. This reveals the non-equivalence between time reversibility and reciprocity. The latter is more general: reciprocity can hold for systems in which irreversible processes take place, as a fundamental consequence of Onsager's principle of microscopic reversibility \cite{Casimir}. In the context
of planar chirality, this subtle link plays a fundamental role, as it will be discussed in section {\it section}.
\section{Optical chirality}
\subsection{Chiral Jones Matrix }
\indent Following the operational definition of Lord Kelvin, the study of chirality demands to characterize the
optical behavior of the considered system through a planar mirror symmetry
$\Pi_{\vartheta}$. By definition, an in-plane symmetry axis making
an angle $\vartheta/2$ with respect to the $x$-direction is associated with
transformation matrices
\begin{eqnarray} {\Pi}_{\vartheta}=\left(\begin{array}{cc}
\cos{\vartheta}& \sin{\vartheta}
\\ \sin{\vartheta}& -\cos{\vartheta}
\end{array}\right), & \overline{{\Pi}_{\vartheta}}=\left(\begin{array}{cc}
0 & e^{-i\vartheta}
\\ e^{+i\vartheta}& 0
\end{array}\right),\label{pi}
\end{eqnarray}
respectively written in cartesian and circular bases with ${\Pi}_{\vartheta=0}={\Pi}_{x}$. Through that $\Pi_{\vartheta}$ symmetry operation, the Jones matrix transforms as $\hat{J}_{\Pi}=\hat{\Pi}_{\vartheta}\hat{J}\hat{\Pi}_{\vartheta}^{-1}$ in the cartesian basis and as
\begin{eqnarray}
\overline{{J}}_{\Pi}=\overline{{\Pi}_{\vartheta}}\cdot\overline{{J}}\cdot\overline{{\Pi}_{\vartheta}}^{-1}=\left(\begin{array}{cc}
J_{rr}& J_{rl}e^{-i2\vartheta}
\\ J_{lr}e^{+i2\vartheta}& J_{ll}
\end{array}\right).\label{mirror}
\end{eqnarray}
in the circular basis. \\
\indent With Kelvin's definition, a system will be optically non-chiral if, and only if, it is invariant under $\Pi_{\vartheta}$, meaning that ${J}={J}_{\Pi}$ or equivalently that the operators respectively associated with the Jones matrix and the mirror-symmetry matrix commute as $[\hat{J},\hat{\Pi}_{\vartheta}]=\hat{J}\hat{\Pi_{\vartheta}}-\hat{\Pi}_{\vartheta}\hat{J}=0$.
The invariance condition $\overline{{J}}=\overline{{J}}_{\textrm{mirror}}$ enforces two constraints on the Jones matrix coefficients, namely that
\begin{eqnarray}
J_{ll}=J_{rr} \ \ {\rm and} \ \ J_{rl}=J_{lr}e^{2i\vartheta}.  \label{cond}
\end{eqnarray}
This implies that the Jones matrix associated with a non-chiral optical system has the following general form
\begin{eqnarray}
{J}_{\textrm{mirror}}=\left(\begin{array}{cc} A+B\cos{\vartheta}&
B\sin{\vartheta}
\\ B\sin{\vartheta}& A-B\cos{\vartheta}
\end{array}\right) \label{eq18}
\end{eqnarray}
\indent By contrapositive of conditions (\ref{cond}), we see that
\begin{eqnarray*}
 \textrm{\underline{Theorem:}}  \nonumber \\
 && \emph{Optical chirality is possible if, and only if,} \nonumber  \\
&& J_{ll}\neq J_{rr}  \  \textrm{OR} \ |J_{lr}|\neq |J_{rl}|, 
\end{eqnarray*}
OR being the logical disjunction. \\
\indent This constitutes a theorem equivalent to Kelvin's statement that an optically chiral system has no mirror symmetry, with $J\neq J_{\Pi}$ or, equivalently, with non-commuting operators respectively associated with the Jones matrix and the mirror-symmetry matrix as
\begin{eqnarray}
[\hat{J},\hat{\Pi}_{\vartheta}]=\hat{J}\hat{\Pi_{\vartheta}}-\hat{\Pi}_{\vartheta}\hat{J} \neq 0 \textrm{ for any $\vartheta$.} \label{eq19}
\end{eqnarray}
Such a Jones matrix can be written in the following form:
\begin{eqnarray}
\overline{{J}}=\left(\begin{array}{cc} (J_{ll}+J_{rr})/2& J_{lr}
\\ J_{rl}& (J_{ll}+J_{rr})/2
\end{array}\right)
+\frac{(J_{ll}-J_{rr})}{2}\left(\begin{array}{cc} 1& 0
\\ 0& -1
\end{array}\right).\nonumber \\
\label{eq20}
\end{eqnarray}
\indent Additionally for an optically chiral system, the application of a
mirror symmetry ${\Pi}_{\vartheta}$ provides new chiral optical
structures called enantiomers which are characterized by their Jones
matrices $\overline{{J}}^{\text{enant.}}(\vartheta)$. By definition
we have
\begin{eqnarray}
\overline{{J}}^{\text{enant.}}(\vartheta)=\overline{{\Pi}_{\vartheta}}\cdot\overline{{J}}\cdot\overline{{\Pi}_{\vartheta}}^{-1}\neq\overline{{J}}.
\label{enatiomer}
\end{eqnarray}
In general, the lack of rotational invariance of $\overline{{J}}$
implies that these enantiomorphic matrices depend specifically on
the mirror reflection $\overline{{\Pi}_{\vartheta}}$ chosen in the
definition given by Eq.~(\ref{eq20}). Thus, there are actually infinite
numbers of such enantiomers.\\
\indent Three important classes of chiral systems can be derived from the truth table associated with the theorem:\\
i) A first class satisfying $J_{ll}\neq J_{rr}$ but with
$|J_{lr}|= |J_{rl}|$. For reasons presented below, this class will be named the ``optical activity class'' or
${E}_{\textrm{o.a.}}$. This class is, until recently, the class essentially discussed in the literature.\\
ii) A second class corresponding to $J_{ll}= J_{rr}$ but satisfying the
constraint $|J_{lr}|\neq |J_{rl}|$. This class will be named in the
following the ``(genuine) planar chirality class'' ${E}_{\textrm{2D}}$, for
reasons also to be given further down. \\
iii) A third class associated with $J_{ll}\neq J_{rr}$ and $|J_{lr}|\neq |J_{rl}|$. This is the most general class of optically chiral system coined as the ``optical chirality class'' and which will be described below in details. \\
\indent Our point is that these chiral classes correspond to specific spatial relations of chiral systems with respect to 3D space. Such relations are fundamental to the characterization of chiral objects, which depends on the shape of the objects and the dimension of the space within which the objects are probed  \cite{Arnaut}. In optics, these relations can be unveiled through reciprocity: as light propagates in 3D space, the effect of optical path reversal through any chiral object will reveal its relation with surrounding space. Analyzing the behavior of the objects concerning reciprocity allow characterizing the relation of any chiral objects with respect to 3D space, from which a classification of the chirality type can be drawn.

\subsection{Optical activity}
\indent  One of the most illustrating example of geometrical
chirality in nature is the helix. The helix is intimately linked
to the most know form of optical chirality namely optical activity
or natural gyrotropy. For example, several natural systems like
sugar molecules and quartz crystal possess a helicoidal structure
and show indeed optical activity properties such as rotatory power
i.e. circular birefringence, or circular dichroism (i.e. a
differential absorbtion for RCP and LCP light). \\
\indent For the present
purpose one of the most relevant property of helices concerns their
sense of ``twist''. It is indeed a basic fact that the twist orientation
of an helix with its axis along Z is invariant through the
rotation ${R}_{X}$: such an helix looks actually quite the same
when watched by an observer in the $+Z$ or $-Z$ direction. This is mathematically rooted in the fact that the helix is a 3D object observed in a 3D space.   \\
\indent A particular application of the geometrical analogy is the case of
an isotropic and homogenous distribution of helices which is
indeed an extreme limit in which the system cannot be physically
(in particular) distinguished when watched from the front or the
back side. This is the case of the sugar molecules solution
considered by Arago and Pasteur in their pioneer works on optical
activity~\cite{Arago,Pasteur}. However, it would be an
oversimplification to limit optical activity to such totally
invariant system since in general even an helix with its axis
oriented along $Z$ is not completely invariant through ${R}_{X}$
(although the sense of twist obviously is). Indeed, due to its
finite length one will have after application of ${R}_X$ in
general to rotate the helix by a given supplementary angle
$\vartheta$ around $Z$ in order to return to the original helix
(as seen from the front side). The analogy with the
helix will give us a simple way to generalize our discussion and to define a criteria for optical activity.\\
\indent More precisely, in the limit of the paraxial approximation
considered here the question we should ask to ourself is what must
be the precise structure of the Jones matrix $J$ if we impose that
${J}^{\textrm{flip}}$ (see Eq.~(\ref{flip})) is, up to a rotation
${R}_{\vartheta}$ by an angle $\vartheta$ around the $Z$ axis,
identical
to ${J}$?\\
This last condition reads actually in the cartesian basis
\begin{eqnarray}
{{J}}^{\textrm{flip}}={{\Pi_x}}\cdot{{J}^{\textrm{T}}}
\cdot{{\Pi_x}}^{-1}={{R}_{\vartheta}}\cdot {{J}}\cdot
{{R}_{\vartheta}}^{-1}. \label{eq22}
\end{eqnarray}
It is preferable to use the  $\mathbf{L},\mathbf{R}$ basis and the
previous condition becomes
\begin{eqnarray}
\overline{{J}}^{\textrm{T}}=\overline{{R}_{\vartheta}}\cdot
\overline{{J}}\cdot \overline{{R}_{\vartheta}}^{-1}.
\end{eqnarray}
The rotation matrix is defined by \begin{eqnarray}
{R}_{\vartheta}=\left(\begin{array}{cc} \cos{\vartheta}&
\sin{\vartheta}
\\ -\sin{\vartheta}& \cos{\vartheta}
\end{array}\right), & \overline{{R}_{\vartheta}}=\left(\begin{array}{cc}
 e^{+i\vartheta} &0
\\0& e^{-i\vartheta}
\end{array}\right).\label{eq24}
\end{eqnarray} 
From equations (\ref{flip},\ref{mirror},\ref{eq22}), we
deduce directly that the previous condition imposes
$J_{lr}e^{i2\vartheta}=J_{rl}$ (i.e., $|J_{lr}|=|J_{rl}|$).
Therefore the Jones matrix  takes the following form
\begin{eqnarray}
{J}=\left(\begin{array}{cc} J_{ll}& J_{lr}
\\ J_{lr}e^{+i2\vartheta}& J_{rr}
\end{array}\right).\end{eqnarray}
By comparing with the condition for the absence of mirror symmetry
(see Eq.~(\ref{eq19})) and our theorem  we see that this $J$ matrix is chiral
if and only if $J_{ll}\neq J_{rr}$. The class of all the matrices
\begin{eqnarray}
\overline{{J}}_{\textrm{o.a.}}=\left(\begin{array}{cc} J_{{ll}}&
J_{{lr}}
\\ J_{{lr}}e^{+i2\vartheta}& J_{{rr}}
\end{array}\right),\label{OA}
\end{eqnarray}
fulfilling these conditions is physically associated with the
phenomenon of optical activity. This justifies the name given to
${E}_{\textrm{o.a.}}$. Equivalently stated this result means that
Eq.~(\ref{OA}) defines the most general Jones matrices which are i)
chiral and ii) such that the optical signature of chirality is, up
to a rotation ${{R}_{\vartheta}}$, invariant after reversal of the
direction of propagation through the system. Clearly, reciprocity
here dictates the rules. Importantly, Eq.~(\ref{OA}) can also be
written
\begin{eqnarray}
\overline{{J}}_{\textrm{o.a.}}=\left(\begin{array}{cc}
(J_{ll}+J_{rr})/2& J_{lr}
\\ J_{lr}e^{+i2\vartheta}& (J_{ll}+J_{rr})/2
\end{array}\right)
+\frac{J_{ll}-J_{rr}}{2}\left(\begin{array}{cc} 1& 0
\\0& -1
\end{array}\right),\nonumber \\
\end{eqnarray} that is as the sum of a matrix
$\overline{{J}}_{\textrm{mirror}}$ obeying Eqs.~(\ref{cond},\ref{eq18}) (i.e., having
an in-plane mirror symmetry axis) and of a matrix
$\left(\begin{array}{cc} \delta & 0
\\0& -\delta
\end{array}\right)$ (with $\delta\neq0$) which actually induces the
chiral behavior. In the cartesian basis we can equivalently write
\begin{eqnarray}
{J}_{\textrm{o.a.}}=\left(\begin{array}{cc} A+B\cos{\vartheta}&
B\sin{\vartheta}
\\ B\sin{\vartheta}& A-B\cos{\vartheta}
\end{array}\right)
+\left(\begin{array}{cc} 0& i\gamma
\\ -i\gamma& 0
\end{array}\right),\end{eqnarray}
with $A=(J_{ll}+J_{rr})/2$, $B=J_{lr}e^{i\vartheta}$ and $\gamma=
(J_{rr}-J_{ll})/2$. The presence of the antisymmetrical part is the
signature of chirality and the coefficient $\gamma\neq0$ is called
(natural) gyromagnetic factor.\\
\indent An important particular case concern Jones matrices which
are invariant through a rotation by an angle $\vartheta$ around the
$Z$ axis. Such a rotation is defined by the matrix ${R}_{\vartheta}$
with
\begin{eqnarray} {R}_{\vartheta}=\left(\begin{array}{cc}
\cos{\vartheta}& \sin{\vartheta}
\\ -\sin{\vartheta}& \cos{\vartheta}
\end{array}\right), & \overline{{R}_{\vartheta}}=\left(\begin{array}{cc}
 e^{+i\vartheta} &0
\\0& e^{-i\vartheta}
\end{array}\right).\label{eq29}
\end{eqnarray}
The invariance by rotation implies that the matrix
$\overline{{J}}_{R}=\overline{{R}_{\vartheta}}\cdot\overline{{J}}\cdot\overline{{R}_{\vartheta}}^{-1}$
equals $\overline{{J}}$, that is:
\begin{eqnarray}
\overline{{J}}_{\textrm{rotation axis}}=\left(\begin{array}{cc}
J_{ll}& 0
\\ 0& J_{rr}
\end{array}\right).\label{rotation}
\end{eqnarray} If $J_{ll}\neq J_{rr}$ then Eq.~(\ref{rotation}) is clearly a
particular case of Eq.~(\ref{OA}) which actually describes optical
activity in isotropic media, such as quartz crystals or molecular
solutions, and corresponds to the circular birefringence (and
dichroism) introduced by Fresnel in 1825. It is interesting to
observe that this is also the matrix which is associated with the
gammadions artificial structure considered in
~\cite{Papakostas,Schwanecke,Vallius,Gonokami} and which have a
four-fold rotational invariance around $Z$.\\
\indent Following its definition, the Jones enantiomorphic matrix
associated with
$\overline{{J}}_{\textrm{o.a.}}=\left(\begin{array}{cc} A& 0
\\ 0& B
\end{array}\right)$ writes as $\overline{{J}}^{\text{enant.}}_{\textrm{o.a.}}(\vartheta)=\left(\begin{array}{cc} B&
0
\\ 0& A
\end{array}\right)$. Because of rotational invariance,
$\overline{{J}}^{\text{enant.}}_{\textrm{o.a.}}(\vartheta)$ is
independent of $\vartheta$. These two enantiomorphic matrices are
associated with opposite optical rotatory powers.\\
\indent Consider for example the Jones matrix associated with
optical activity in an isotropic medium, such as a random
distribution of helices for example. From
Eqs.~(\ref{flip},\ref{enatiomer},\ref{rotation}) it is immediately
seen that $\overline{{J}}^{\textrm{flip.}}=\overline{{J}}$. This
invariance means that an observer illuminating such a system cannot
distinguish the two sides from one other. This well known property
explains in particular why an optically active medium cannot be used
as an optical isolator: reciprocity prohibits such a scenario. In
this context we point out that nothing here forbid unitarity to
hold. In the particular case of a Jones matrix represented by a
rotation (see Eq.~(\ref{eq24})) we have indeed $J^{-1}=J^{\dagger}$. We will
see in the next section that this is not the case for 2D chirality
where losses are unavoidable. \\
\indent It is finally useful to remark that the ensemble of all the
matrices $\overline{{J}}_{\textrm{o.a.}}$, i.e.,
${E}_{\textrm{o.a.}}$, is not closed for the addition and the
product of matrices (that is the sum or the product of chiral matrix
belonging to ${E}_{\textrm{o.a.}}$ is not necessary contained in
${E}_{\textrm{o.a.}}$). For example by combining two enantiomers
characterized by the matrices $\left(\begin{array}{cc} A & 0
\\0& B
\end{array}\right)$ and $\left(\begin{array}{cc} B & 0
\\0& A
\end{array}\right)$ we get
\begin{eqnarray}
\left(\begin{array}{cc} A & 0
\\0& B
\end{array}\right)+\left(\begin{array}{cc} B& 0
\\0& A
\end{array}\right)=(A+B)\left(\begin{array}{cc} 1 & 0
\\0& 1
\end{array}\right),\nonumber\\
\textrm{and,} \left(\begin{array}{cc} A & 0
\\0& B
\end{array}\right)\cdot\left(\begin{array}{cc} B& 0
\\0& A
\end{array}\right)=(A.B)\left(\begin{array}{cc} 1 & 0
\\0& 1
\end{array}\right),
\end{eqnarray} which are obviously not chiral and correspond to
what is called a racemic medium (that is a mixture or a
juxtaposition of opposite enantiomers).\\
We point out that the
rotation matrices considered in the present discussion involve only
an axis of rotation oriented along the $z$ direction.  It could be
interesting to consider more general 3D rotation with an axis
arbitrarily oriented. However, the case of planar chirality to be
considered in the next section would thus be problematic in our
classification since as stated in the introduction  a planar chiral
structure can be brought into congruence with its mirror image if it
is lifted from the plane. The classification used here which
consider explicitly as distinct the two ensembles
${E}_{\textrm{2D}}$ and ${E}_{\textrm{o.a.}}$ will appear actually
very convenient since  (as shown in section 3.4) any  chiral matrix
can be split into a first matrix belonging to ${E}_{\textrm{2D}}$
and a second matrix belonging to ${E}_{\textrm{o.a.}}$.\\
\subsection{Planar chirality}
\indent Despite its fundamental importance the previous analysis of
optical activity does not exhaust the problem of chirality in optics. As we wrote in the introduction, 2D chirality characterizes, by definition, a system \emph{which
can not be put into congruence with itself until left from the
plane}  (for a more mathematical discussion see
\cite{Arnaut,Osipov}). This corresponds to a different chirality class than optical activity, as it can be simply seen. If instead of a 3D helix one considers a 2D spiral contained within the $(x,y)$
plane, one has obviously a system that has a dimension lower than the dimension of its surrounding space. A flip of the structure is now possible, when it was not for the helix. As we discussed above, this discussion corresponds optically to a change of twist orientation when the light path is reversed, and clearly motivates the experimental demonstration for this second
class of chiral objects which are planar.  \\
\indent Two geometrical examples of such planar
chiral object are the Archimedean spiral which has not point
symmetry and the gammadion which possesses  rotational invariance.
Since optically gammadion are associated  with gyrotropy i.e.
essentially 3D effect as discussed in the previous section we can
already think that  such gammadions are not genuine 2D chiral object
from the optical point of view (even though it is obviously the case
from basic geometrical considerations).  We will  here consider more
in details this 2D chirality class of
Jones matrices, i.e, ${E}_{\textrm{2D}}$.\\
The planar chirality class ${E}_{\textrm{2D}}$
 was until very recently completely ignored in
the literature and concerns chiral systems characterized by the
conditions $J_{ll}=J_{rr}$ and a Jones matrix of the form:
\begin{eqnarray}
\overline{{J}}_{\textrm{2D}}=\left(\begin{array}{cc} J_{ll}& J_{lr}
\\ J_{rl}& J_{ll}
\end{array}\right)\textrm{ with $|J_{lr}|\neq
|J_{rl}|$.} \label{eq32}
\end{eqnarray}
\indent The condition
$|J_{lr}|\neq |J_{rl}|$ actually leads to chirality. As we will show below, the equality condition on diagonal elements correspond to reciprocity. \\
\indent It is also important to observe that, since Eq.~(\ref{eq32}) is
different from Eqs.~(\ref{cond},\ref{rotation}), the matrix
$\overline{{J}}_{\textrm{2D}}$ not only has no mirror symmetries,
but it has additionally no rotational invariance. This means, that
$\overline{{J}}_{\textrm{2D}}$ can only be associated with chiral
systems without any point symmetries, such as for example an
Archimedean spiral or a fish-scale structure. A gammadion structure
with its four-fold rotational invariance can not display such
optical property. It should also be remarked that the fish-scale
structure considered in Ref.~\cite{Fedotov} actually has a central
point symmetry, i.e., a two-fold rotation axis. However from the
point of view of Jones matrix such transformation is equivalent to
the identity, as it is immediately seen by writing $\vartheta=\pi$
in Eq.~(\ref{eq29}), and consequently the structure of $\overline{{J}}$ is not
constrained by such transformation.\\
\indent Same as for $\overline{{J}}_{\textrm{o.a.}}$ one can define
enantiomers structures by the relation
\begin{eqnarray}
\overline{{J}}^{\text{enant.}}_{\textrm{2D}}=\overline{{\Pi}_{\vartheta}}\cdot\overline{{J}}_{\textrm{2D}}\cdot\overline{{\Pi}_{\vartheta}}^{-1}=\left(\begin{array}{cc}
J_{ll}& J_{rl}e^{-2i\vartheta}
\\ J_{lr}e^{2i\vartheta}& J_{ll}
\end{array}\right).\label{enant2d}
\end{eqnarray}
$\overline{{J}}^{\text{enant.}}_{\textrm{2D}}$ of course belongs to ${E}_{\textrm{2D}}$.\\
\indent We will now go back to the reciprocity theorem  and consider
the properties of planar chiral system from the point of view of
path reversal. Same as for optical activity the geometrical analogy
appears  very convenient for characterizing the reciprocal
properties of chiral planar systems.  The archetype of planar chiral
objects is, as we already mentioned,   the Archimedean spiral with
no point symmetry.   Watching such a spiral from one side or the
other changes obviously the sense of twist.  This contrasts strongly
with the case of the helix of of gammadion discussed before. This
suggests the following definition:  the chiral system characterized
by the the Jones matrix  $J$ is plan chiral if and only if Eq.~(\ref{eq19}) is
satisfied and if
\begin{eqnarray}
{{J}}^{\textrm{flip}}={{\Pi_x}}\cdot{{J}^{\textrm{T}}}
\cdot{{\Pi_x}}={{R}_{\vartheta}}\cdot
({{\Pi_x}}\cdot{{J}}\cdot{{\Pi_x}})\cdot {{R}_{\vartheta}}^{-1}.\label{eq34}
\end{eqnarray} where ${{\Pi_x}}\cdot{{J}}\cdot{{\Pi_x}}$
corresponds to a particular enantiomorphic  Jones matrix of $J$ (see
Eq.~(\ref{enant2d})) parameterized by the angle $\vartheta$.\\
This condition is in the $\mathbf{L},\mathbf{R}$ basis equivalent to
\begin{eqnarray}
\left(\begin{array}{cc} J_{ll}& J_{rl}
\\ J_{lr}& J_{rr}
\end{array}\right)=\left(\begin{array}{cc} J_{rr}& J_{rl}e^{2i\vartheta}
\\ J_{lr}e^{-2i\vartheta}& J_{ll}
\end{array}\right),
\end{eqnarray} which admits a solution if and only if $J_{ll}=J_{rr}$ and
$\vartheta=0$ or $\pi$. Since ${{R}_{0}}=- {{R}_{\pi}}=I$ (i.e., the
identity operator) it means that the rotation plays here no role in
the definition and that we could have reduced our reasoning to the
condition\begin{eqnarray}
{{J}}^{\textrm{flip}}={{\Pi_x}}\cdot{{J}}\cdot{{\Pi_x}} \textrm{
,i.e., } {{J}}^{T}={{J}}. \label{eq36}
\end{eqnarray}
Additionally, in order to satisfy Eq.~(\ref{eq19}) we must necessarily have
$|J_{lr}|\neq|J_{rl}|$. This is rigorously equivalent to the
definition of the class ${E}_{\textrm{2D}}$ given above.\\
In the cartesian basis this means
\begin{eqnarray}
{{J}}_{\textrm{2D}}=\left(\begin{array}{cc} \varepsilon_+& \Gamma
\\ \Gamma& \varepsilon_-
\end{array}\right)={{J}}_{\textrm{2D}}^T \label{eq37}
\end{eqnarray} with $\varepsilon_\pm=J_{ll}\pm(J_{lr}+J_{rl})/2$, and
$\Gamma=i(J_{lr}-J_{rl})/2$. The condition for chirality
$|J_{lr}|\neq|J_{rl}|$ implies $\varepsilon_+\neq \varepsilon_-$ and
$\Gamma\neq0$.  This condition for chirality also implies a stronger
restriction:\\
Indeed, writing the non-diagonal coefficients $J_{lr}$,$J_{rl}$ in
the polar form $J_{lr}=ae^{i\phi}$ and $J_{rl}=be^{i\chi}$ (with
$a$, $b$ the norms and $\phi$, $\chi$ the phases) we can define the
ratio
$$\eta=\frac{\Gamma}{(\varepsilon_+-\varepsilon_-)}=i\frac{J_{lr}-J_{rl}}{J_{lr}+J_{rl}}$$
which thus becomes
\begin{eqnarray}
\eta=i\frac{1-\frac{b^2}{a^2}}{1+\frac{b^2}{a^2}}-2\frac{\frac{b}{a}}{1+\frac{b^2}{a^2}}\sin{(\phi-\chi)}.
\end{eqnarray} The condition for chirality therefore implies:
\begin{eqnarray}
\textrm{Imag}[\frac{\Gamma}{(\varepsilon_+-\varepsilon_-)}]\neq 0. \label{eq39}
\end{eqnarray}
\indent In the context of planar chirality it is also useful to
check if ${E}_{\textrm{2D}}$ is closed with respect to the matrices
addition and multiplication. Same as for ${E}_{\textrm{o.a.}}$, it
is obvious that this is not the case since by summing
\begin{eqnarray}\overline{{J}}_{\textrm{2D}}=\left(\begin{array}{cc} \alpha &
\beta
\\ \gamma& \alpha
\end{array}\right)\label{eq40}
\end{eqnarray} and
\begin{eqnarray}\overline{{J}}^{\text{enant.}}_{\textrm{2D}}=\left(\begin{array}{cc}
\alpha & \gamma
\\ \beta& \alpha
\end{array}\right)
\end{eqnarray}  one get
\begin{eqnarray}\left(\begin{array}{cc} 2\alpha & \beta+\gamma
\\ \gamma+\beta& 2\alpha
\end{array}\right)\end{eqnarray}  which does not belong to
${E}_{\textrm{2D}}$ (indeed, we have $|J_{lr}|=|J_{rl}|$).
Similarly, for the product of $\overline{{J}}_{\textrm{2D}}$ and
$\overline{{J}}^{\text{enant.}}_{\textrm{2D}}$ one obtain
$\left(\begin{array}{cc} \alpha^2+\beta^2 & \alpha(\gamma+\beta)
\\ \alpha(\gamma+\beta)& \alpha^2+\gamma^2
\end{array}\right)$ which for the same reasons does not also belong to
${E}_{\textrm{2D}}$.\\
Finally, we  point out that  all genuine  2D chiral systems share
an other important property namely a breaking of time invariance
or reversal~\cite{Fedotov,Fedotov2,Fedotov3,Drezet1}. To
understand this peculiarity we go back to our discussion of time
reversal and consider the condition that the Jones matrix
$\overline{J}_{2D}$ defined by Eq.~(\ref{eq32}) should fulfill in order to
be unitary, i.e.,
$\overline{J}_{2D}^{-1}=\overline{J}_{2D}^\dagger$. From Eq.~(\ref{eq40})
this leads to the relation:
\begin{eqnarray}\frac{1}{\alpha^2-\beta\gamma}\left(\begin{array}{cc} \alpha &
-\beta
\\ -\gamma& \alpha
\end{array}\right)=\left(\begin{array}{cc} \alpha^\ast &
\gamma^\ast
\\ \beta^\ast& \alpha^\ast
\end{array}\right).\end{eqnarray}  This implies
$\frac{\alpha}{\alpha^2-\beta\gamma}=\alpha^\ast$,
$\frac{\beta}{\alpha^2-\beta\gamma}=-\gamma^\ast$ and
$\frac{\gamma}{\alpha^2-\beta\gamma}=-\beta^\ast$. By taking the
norms of each terms we deduce $|\alpha^2-\beta\gamma|=1$ and
$|\beta|=|\gamma|$. This last equality contradicts the definition
of $\overline{J}_{2D}$ and consequently such a planar chiral Jones
matrix can not be unitary.  This implies that ${J}_{2D}^{\textrm{T}}\neq
{J}_{2D}^{-1,\ast}$ and that therefore ${J}_{2D}^{\textrm{rec.}}$ is
different from ${J}_{2D}^{\textrm{inv.}}$. \\
\indent In other words, a 2D chiral
system provides a perfect illustration that time-reversal is necessary different from reciprocity, i.e.
path reversal. Since time-reversal is a key property of
fundamental physical laws at the {\it microscopic} level, the only
solution is to assume that this breaking of time-reversal at the level of 2D chiral objects is
associated with {\it macroscopic} irreversibility. Indeed, the imaginary
part of the permittivity, for example, is connected to losses and
dissipation into the environment (seen as a thermal bath) and the condition for its
positivity implies a strong irreversibility in the propagation.
Similarly here, 2D optical chirality means that some sources of
irreversibility must be present in order to prohibit unitarity of
the Jones matrix.  This is an interesting example where two
fundamental aspects of nature namely chirality and time
irreversibility (intrinsically linked to the entropic time arrow)
are  intimately connected.
\subsection{Generalization}
\indent The most general Jones matrix $J$ characterizing a chiral
medium can be written:
\begin{eqnarray} \overline{{J}}_{\textrm{chiral}}=\left(\begin{array}{cc}
(J_{ll}+J_{rr})/2& J_{lr}
\\ J_{rl}& (J_{ll}+J_{rr})/2
\end{array}\right)
+\frac{(J_{ll}-J_{rr})}{2}\left(\begin{array}{cc} 1& 0
\\ 0& -1
\end{array}\right),\nonumber \\
\end{eqnarray} where both conditions $J_{ll}\neq J_{rr}$, $|J_{lr}|\neq|J_{rl}|$ are
satisfied. This defines the optical chirality class
${E}_{\textrm{chiral}}$. \\
\indent An important property is that the sum of
a matrix belonging to ${E}_{\textrm{o.a.}}$ with a matrix of
${E}_{\textrm{2D}}$ belongs to ${E}_{\textrm{chiral}}$. To see that
this is obviously the case it is sufficient to remark that the sum
of a matrix $\overline{{J}}_{\textrm{2D}}$ with a matrix
$\overline{{J}}_{\textrm{o.a.}}$ can be written
\begin{eqnarray}
\overline{{J}}_{\textrm{2D}}+\overline{{J}}_{\textrm{o.a.}}=
\overline{{J}}_{\textrm{2D}}+\overline{{J}}_{\textrm{mirror}}+
\left(\begin{array}{cc} \alpha & 0
\\0& -\alpha
\end{array}\right).
\end{eqnarray}
However we also have
\begin{eqnarray}
\overline{{J}}_{\textrm{2D}}+\overline{{J}}_{\textrm{mirror}}=\left(\begin{array}{cc}
\alpha+\alpha'& \beta+\beta'
\\ \gamma+\beta' e^{i\phi}&\alpha+\alpha'
\end{array}\right)\nonumber\\
\end{eqnarray} with $|\beta|\neq|\gamma|$.
This is necessary of the form $\overline{{J}}_{\textrm{2D}}$ since
otherwise we should have $|\beta+\beta'|=|\gamma+\beta'e^{i\phi}|$
in contradiction with the condition $|\beta|\neq|\gamma|$.
Reciprocally any matrices $\overline{{J}}'_{\textrm{2D}}$ can be
written as a sum
$\overline{{J}}_{\textrm{2D}}+\overline{{J}}_{\textrm{mirror}}$
since from the previous result for any
$\overline{{J}}_{\textrm{mirror}}$ the difference
$\overline{{J}}'_{\textrm{2D}}-\overline{{J}}_{\textrm{mirror}}$
belongs to ${E}_{\textrm{2D}}$. This means that a matrix belonging
to ${E}_{\textrm{chiral}}$ can always be written as the sum of a
matrix belonging  to the class  ${E}_{\textrm{o.a}}$ with a matrix
belonging to ${E}_{\textrm{2D}}$. Interestingly, the combination of a spiral and an helix leads to the geometrical shape of the screw. Finally, one can observe that
${E}_{\textrm{chiral}}$ is not close with respect to the matrix
addition and product since it is already not the case for the sub
classes ${E}_{\textrm{o.a.}}$ and ${E}_{\textrm{2D}}$.

\subsection{Eigenstates and chirality: time reversal versus reciprocity}\label{path}
\begin{figure}[h]
\centering\includegraphics[width=8cm]{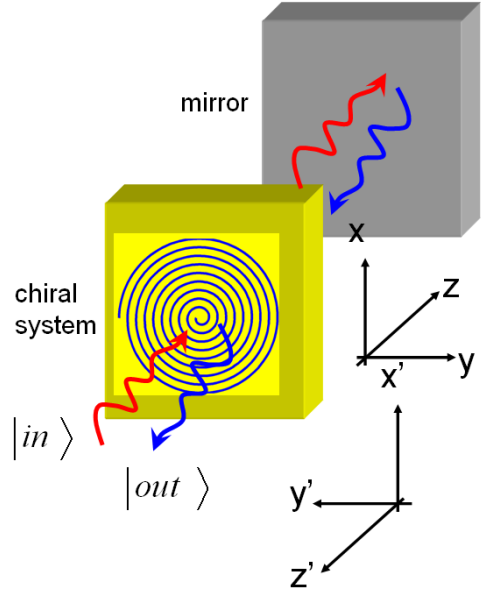}
\caption{Principle of path reversal  by a mirror through a chiral
structure (for example a planar chiral system such as those
considered in Ref.~\cite{Drezet2}). The two coordinate systems
$x,y,z$ and $x'=x,y'=-y,z'=-z$ discussed in the text are
represented.}\label{fig2}
\end{figure}

\indent In the context of reciprocity it is of practical importance
to consider the backward propagation of light through the medium
along the z direction after path reversal by a mirror located after
it
(see Fig.~\ref{fig2}).
This corresponds to the following succession of events: 1) the
initial state, that we write here
$|in\rangle=E^{\textrm{(in)}}_x|x\rangle+E^{\textrm{(in)}}_y|y\rangle=E^{\textrm{(in)}}_L|L\rangle+E^{\textrm{(in)}}_R|R\rangle$
(instead of $\mathbf{E}^{\textrm{(in)}}=E^{\textrm{(in)}}_x
\hat{\mathbf{x}}+ E^{\textrm{(in)}}_y \hat{\mathbf{y}}$ used in
section \ref{Lorentz}), propagates through the chiral medium and we obtain
afterward the new state $|2\rangle=\hat{J}|in\rangle$. 2) The
reflection by the mirror induces a change in the electric field sign
and also reverses the path propagation direction. By rotating the
coordinate axes by an angle $\pi$ around the $x$ axis we preserve
(as explained before) the handedness of such coordinate system as
well as the positive sign of the propagation direction. This means,
that in this new basis the vector $|2\rangle$ evolves as
$|3\rangle=-\hat{\Pi}_x|2\rangle$. 3) The backward propagation
through the medium  is described by the 'flip' operator
$\hat{J}^{\textrm{flip}}$ and the final state (in the new coordinate
system) reads $|out\rangle=\hat{J}^{\textrm{flip}}|3\rangle$. We
have consequently
\begin{eqnarray}
|out\rangle=-\hat{J}^{\textrm{flip}}\hat{\Pi}_x\hat{J}|in\rangle=-\hat{J}^{\textrm{flip}}\hat{J}^{\textrm{enant.}}\hat{\Pi}_x|in\rangle\label{back}
\end{eqnarray}
where  by definition
$\hat{J}^{\textrm{enant.}}=\hat{\Pi}_x\hat{J}\hat{\Pi}_x$.\\
\indent In order to analyze  the effect of Eq.~(\ref{back}) we will
first study more in details the eigenstates and eingenvalues  of the
chiral Jones matrix discussed in this chapter. First, we consider
the case of optical activity where
\begin{eqnarray}
\overline{{J}}_{\textrm{o.a.}}=\left(\begin{array}{cc} \alpha & 0
\\ 0& \beta
\end{array}\right)=\overline{{J}}^{\textrm{flip}}.\end{eqnarray} 
The
two eigenstates are $|L\rangle$ and $|R\rangle$ corresponding to
eigenvalues $\alpha$ and $\beta$ respectively. However,
$\hat{J}^{\textrm{flip}}=\hat{J}\neq\hat{J}^{\textrm{enant.}}$ and
while the two eigenstates of $\hat{J}^{\textrm{enant.}}$ are still
$|L\rangle$ and $|R\rangle$ the eigenvalues are now exchanged i.e.
$\beta$ and $\alpha$ respectively. Therefore, a direct application
of Eq.~(\ref{back}) on any vector $|in\rangle=a|L\rangle+b|R\rangle$
lead to
\begin{eqnarray}
|out\rangle=-\hat{J}\hat{J}^{\textrm{enant.}}\hat{\Pi}_x|in\rangle=-\alpha\beta(a|R\rangle+b|L\rangle).
\end{eqnarray}
If we are only interested into the field expression in the original
coordinate system we can alternatively rewrite \begin{eqnarray}
\hat{\Pi}_x|out\rangle=-\alpha\beta(a|L\rangle+b|R\rangle)=-\alpha\beta|in\rangle.
\end{eqnarray} This is a direct formulation of the fact that path
reversal should lead us here back to the initial state $|in\rangle$
as expected.  It illustrates the impact of reciprocity on
propagation and show that a 3D chiral medium  doesn't act as an
optical isolator. We also point out that for a loss-less ideal
medium represented by a unitary rotation matrix
$\overline{{J}}_{\textrm{o.a.}}=\overline{{R}_{\vartheta}}$ with
eigenvalues $e^{\pm i\vartheta}$  we have exactly
$\hat{\Pi}_x|out\rangle=-|in\rangle$ which, up to the minus sign
coming from the mirror reflection, is a perfect illustration of time
reversal and symmetry for natural optical activity. \\
\indent We now consider 2D planar chirality. The eigenstates and
eingenvalues of the chiral Jones matrix
\begin{eqnarray}
\overline{{J}}_{\textrm{2D}}=\left(\begin{array}{cc} \alpha & \beta
\\ \gamma& \alpha
\end{array}\right)\end{eqnarray} are by definition states $|\pm\rangle$
defined by
$\hat{J}_{\textrm{2D}}|\pm\rangle=\lambda_{\pm}|\pm\rangle$. After
straightforward calculations we obtain
\begin{equation}
\lambda_{\pm}=\alpha\pm\sqrt{(\beta\gamma)},
\end{equation} and
\begin{equation}
|\pm\rangle=\frac{\sqrt{\beta}|L\rangle\pm\sqrt{\gamma}|R\rangle}{|\beta|+|\gamma|}.
\end{equation}
Using similar methods we can easily find eigenstates and values of
the reciprocal matrix
$\overline{{J}}_{\textrm{2D}}^{\textrm{flip}}=\overline{{J}}^{\text{enant.}}_{\textrm{2D}}$
such as
$\hat{J}_{\textrm{2D}}^{\textrm{flip}}|\pm\rangle_{\text{enant.}}=\lambda'_{\pm}|\pm\rangle_{\text{enant.}}$.
The eigenvalues are the same as for $\overline{{J}}_{\textrm{2D}}$
,i.e., $\lambda'_{\pm}=\lambda_{\pm}$ but the eigenvectors are now
\begin{equation}
|\pm\rangle_{\text{enant.}}=\frac{\sqrt{\gamma}|L\rangle\pm\sqrt{\beta}|R\rangle}{|\beta|+|\gamma|}.
\end{equation}
Importantly here $\hat{J}^{\textrm{flip}}=\hat{J}^{\textrm{enant.}}$
and therefore by applying Eq.~(\ref{back}) on the initial states
$|in\rangle=|\pm\rangle$ we get
\begin{eqnarray}
|out\rangle=-(\hat{J}^{\textrm{flip}})^2\hat{\Pi}_x|in\rangle=\mp\lambda_{\pm}^2|\pm\rangle_{\text{enant.}}
\end{eqnarray} where we used the relations
$\hat{\Pi}_x|L\rangle=|R\rangle$, $\hat{\Pi}_x|R\rangle=|L\rangle$,
and $\hat{\Pi}_x|\pm\rangle=\pm|\pm\rangle_{\text{enant.}}$. Like we
did for optical activity  we can go back to the initial coordinate
system $x,y,z$ and the final states read now
\begin{eqnarray}
\hat{\Pi}_x|out\rangle=-\lambda_{\pm}^2|\pm\rangle
\end{eqnarray} As before that again illustrates the effectiveness of the reciprocity
principle. Furthermore, since the transformation is not unitary we
could not obtain such a result using $J^{\textrm{inv.}}$ as defined
by Eq.~(\ref{eq14}).
\section{Discussion and examples}
\indent As we mentioned in the introduction it is very interesting
to observe that the property concerning the change of twist for
genuine 2D chiral systems when watched from two different sides
stirred a considerable debates in the recent year in the context
of metamaterials. \indent To understand this more in details we
remind that partly boosted by practical motivations, such as the
quest of negative refractive lenses~\cite{Pendry} or the
possibility to obtain giant optical activity for applications in
optoelectronics, there is currently a renewed
interest~\cite{Pendry,Papakostas,Schwanecke,Vallius,Gonokami,Canfield,Canfield2,Zhang,Decker,Plum,Rogacheva}
in the optical activity in artificial photonic media with planar
chiral structures. It was shown for instance that planar
gammadionic structures, which have by definition no axis of
reflection but a four-fold rotational
invariance~\cite{Papakostas,Vallius}, can generate optical
activity with giant gyrotropic
factors~\cite{Gonokami,Rogacheva,Plum,Decker}. Importantly, and in
contrast to the usual three dimensional (3D) chiral medium (like
quartz and its helicoidal structure~\cite{Hecht,Bose}), planar
chiral structures change their observed handedness when the
direction of light is reversed through the
system~\cite{Papakostas,Barron1}. This paradoxically challenged
Lorentz principle of reciprocity~\cite{Landau} (which is known to
hold for any linear non magneto-optical media) and stirred up
considerable debate~\cite{Papakostas,Schwanecke,Gonokami,Barron2}
which came to the conclusion that optical activity cannot be a
purely 2D effect and always requires a small dissymmetry between
the
two sides of the system~\cite{Gonokami,Rogacheva,Plum,Decker}.\\
\indent This point becomes more clear from the previous definitions and
discussion concerning chirality and reciprocity. Indeed, a gammadion
being rotationally invariant its optical properties must be
characterized by a Jones matrix belonging to ${E}_{\textrm{o.a.}}$
i.e.,
\begin{eqnarray}\overline{{J}}_{\textrm{o.a.}}=\left(\begin{array}{cc} A& 0
\\ 0& B
\end{array}\right),\end{eqnarray} with $A\neq B$
However since geometrically the 2D gammadion change its sense of
twist when watched from the other side the discussion concerning
reciprocity and change of twist developed between Eqs.~(\ref{eq34},\ref{eq39})
imposes that the Jones matrix should also belong to
${E}_{\textrm{2D}}$, i.e.,
\begin{eqnarray}\overline{{J}}_{\textrm{2D}}=\left(\begin{array}{cc} \alpha &
\beta
\\ \gamma& \alpha
\end{array}\right)\end{eqnarray} with  $|\beta|\neq|\gamma|$.
The only possibility for having
$\overline{{J}}_{\textrm{2D}}=\overline{{J}}_{\textrm{o.a.}}$ is to
impose $\gamma=\beta=0$ as well as $A=B=\alpha$ and consequently to
have no optical chiral signature whatsoever. This solves the paradox
and shows that if gammadion generate nevertheless optical activity
with giant gyrotropic factors~\cite{Gonokami,Rogacheva,Plum,Decker}
then the system can not be purely 2D. The third dimension (such as
the presence of substrate for example) is enough to break the
symmetry between the two direction of transmission through the
structure and
there is no violation of reciprocity since the matrix has no anymore reason to satisfy Eq.~(\ref{eq34}) or Eq. (\ref{eq36}) i.e. to belong to ${E}_{\textrm{2D}}$.\\
\indent We also point out that Bunn~\cite{Bunn} and later
L.~Barron~\cite{Barron1,Barron2} already remarked that optical
chirality in 3D and also in 2D (see the next section) characterizes
not only the structure itself but the complete illumination protocol
including the specific orientation of the incident light relatively
to the structure. This idea was recently applied in the context of
metamaterials by Zheludev and coworkers by demonstrating specific
forms of extrinsic optical chirality in which the individual elements
themselves are not chiral while the complete array of such cells is
(due to specific tilts existing
between the incident light and the objects~\cite{PlumPlum}). A related scheme has also been discussed at the level of achiral plasmonic nanohole arrays \cite{MaozNanoLett2012}.\\

\subsection{Surface plasmons and archimedean
spirals: planar chirality gives a twist to light} 
\begin{figure}[h]
\centering\includegraphics[width=10cm]{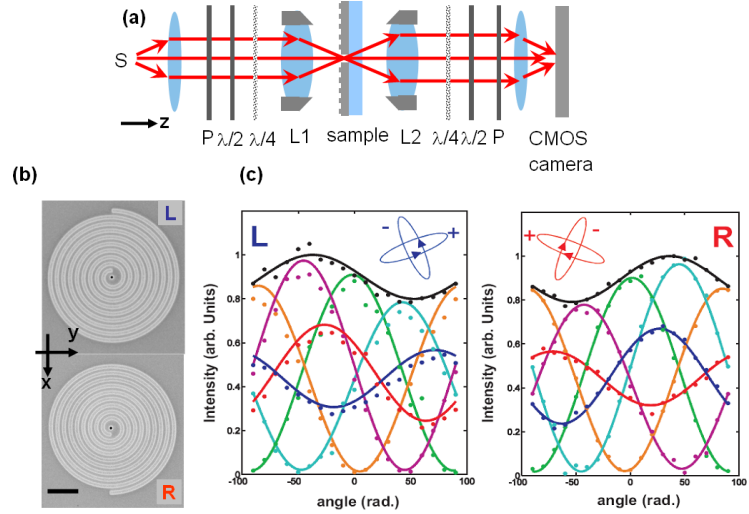} \caption{Chiral
plasmonic metamolecules. (a) Sketch of the polarization tomography
set up used in Refs.~\cite{Drezet1,Drezet2}.(b) Scanning electron
micrographs of the left (L) and right (R) handed enantiomer (mirror
image) planar chiral structures investigated. The scale bar is 3
$\mu$m long. The parameters characterizing the structure are the
following: hole diameter $d=350$ nm, film thickness $h=310$ nm,
grating period $P=760$ nm, groove width $w=370$ nm, and groove depth
$s=80$ nm.(c) Analysis of the polarization states for an input light
with variable linear polarization for both the left (left panel) and
right handed (right panel) individual chiral structures of (a). The
insets show in each panel the ellipses of polarization and the
handedness (arrow) associated with the two co-rotating eingenstates
associated with the Jones matrix $J$ (blue) and
$J^{\textrm{enant.}}$ (red). Details are discussed in the text.
Images adapted from Refs.~\cite{Drezet1,Drezet2}}\label{fig3}
\end{figure}

As mentioned in the introduction the first manifestations
of optical planar chirality were observed by Zheludev and
coworkers~\cite{Fedotov,Fedotov2,Fedotov3} using fish-scale periodic
metal strips on a dielectric substrate. The chiral structures were
first realized at the mm scale for the GHz regime in
2006~\cite{Fedotov} and soon after scaled down to the nanometer
scale  for studied in the near-infrared regime in
2008~\cite{Fedotov3}. Simultaneously with these last studies we
realized planar chiral gratings for surface plasmons on a gold film.
These structure shown in Fig.~\ref{fig3} (b) are Archimede spirals
defined by the parametric equations
\begin{eqnarray}
x(\theta)=\pm\rho(\theta)\cos{(\theta)},&
y(\theta)=\rho(\theta)\sin{(\theta)},
\end{eqnarray}with
\begin{eqnarray}
\rho(\theta)=\frac{P}{2\pi}\theta
\end{eqnarray}
$\theta$ varying between $\theta_{min}=\pi$ and
$\theta_{max}=\theta_{min}+18\pi$. The two possible signs $\pm$
defines two enantiomers (labelled $L$ and $R$ on Fig.~\ref{fig3}(b))
which are reciprocal mirror images obtained after reflection across
the y-z plane $x=0$. Such clock wise or anti clockwise spirals were
milled on a 310 nm thick gold film using focus ion beam methods. For
such Archimede's spirals the length $P\simeq 760$ nm plays obviously
the role of `radial period' since at each increment by an angle
$\delta\theta=2\pi$ the radius increase by an amount $\delta
\rho=P$. The structure looks like the well known ``bull eye's''
circular antennas which are used to resonantly couple monochromatic
light with wavelength $\lambda\simeq P$ impinging normally to the
structure~\cite{Degiron,Lezec}. We point out that $P$ is actually
very closed to the SP wavelength $\lambda_{SP}(\lambda_0)= 760$ nm
which corresponds to the optical wavelength $\lambda_0\simeq 780$
nm~\cite{Drezet1}. However, this small difference
is not relevant here. To increase
this similarity with an usual bull eye antenna the groove depth, and
width were also selected to favor the light coupling to the grating.
Importantly, we also milled a 350 diameter hole centered at the
origin of the spirals (i.e. $x=y=0$) in which light can go through.
Altogether the system acted as a chiral bull eye's antenna focussing
SPs at the center $x=y=0$ where they interfere with the incident
light before being transmitted through the hole thanks to a Fano
like mechanism \cite{Cyriaque1,Cyriaque2,Drezet2,Yuri2}. Spectral
properties of such antennas showed the typical optical resonance
centered at $\lambda\simeq P$ as for their circular or elliptical
cousins~\cite{Genet,Degiron,Lezec,Drezet2}.\\
\indent Chiral optical properties of the two enantiomorphic
structures were studied performing a polarization tomography of the
light transmitted through the hole. The method described in
Ref.~\cite{Drezet1,Drezet2} is based on the experimental
determination of the $4\times4$ Mueller
matrix~\cite{Brehonnet,Genet2}. Such a Mueller matrix $M$
characterizes the polarization transformation applied on the
incident light beam with Stokes vector
\begin{eqnarray}
\textbf{S}^{in}=\left(\begin{array}{c}S_0^{in}\\S_1^{in}\\S_2^{in}\\S_3^{in}\end{array}\right).
\end{eqnarray}
The resultant Stokes vector $\textbf{S}^{out}=M\textbf{S}^{in}$ is
linked to the electric field $E=[E_x,E_y]$ transmitted through the
hole. Furthermore, subsequent theoretical analysis demonstrate a
precise connection between the $2\times 2$ Jones matrix $J$
characterizing the system and the $4\times4$ Mueller matrix
$M$~\cite{Brehonnet}. We used an home made microscope to focus and
control the state of polarization (SOP) of light going through the
 chiral structures. In order to study experimentally the SOP conversion by
our structure, we carried out a complete polarization tomography
using the optical setup sketched in Fig.~\ref{fig3} (a). A laser beam
at $\lambda=780$ nm is focused normally onto the structure by using
an objective L1. The transmitted light is collected by a second
objective L2 forming an Airy spot on the camera  as expected since
the hole behaves like a point source in an opaque gold film. In our
experiments, the intensity is thus defined by taking the maximum of
the Airy spot~\cite{Drezet1,Drezet2}. The SOP of light is prepared
and analyzed with half wave plates, quarter wave
plates, and polarizers located before and after the objectives.\\
\indent To illustrate the polarization conversion induced by the
chiral object on the transmitted light we analyze on Fig.~\ref{fig3}  (c) the
transformation acting on a linearly polarized input light analyzed
after transmission in four orthogonal SOP Stokes vectors along the
directions: $|x\rangle$ (green), $|y\rangle$ (yellow),
$|+45^\circ\rangle$ (cyan), $|-45^\circ\rangle$ (magenta),
$|L\rangle$ (red), and $|R\rangle$ (blue). The total transmitted
intensity is also shown (black). The symmetries between both panel
expected from group theory are observed experimentally in perfect
agreement with the theory discussed in Ref.~\cite{Drezet1}. The
insets show in each panel the ellipses of polarization and the
handedness (arrow) associated with the two corotating eingenstates
associated with the Jones matrix. The good agreement between the
measurements and the theoretical predictions deduced from the Jones
matrices \cite{Drezet1} is clearly seen, together with the mirror
symmetries between the two enantiomers. This agreement shows that
our theoretical hypothesis about the form of the matrices $J$ (see
Eqs.~(\ref{eq37},\ref{eq40}) is experimentally justified. Importantly, the observed
symmetries also imply that for unpolarized light, and in complete
consistency with spectral studies, the total intensity transmitted
by the structures is
independent of the chosen enantiomer. 

\subsection{Chiral surface plasmons and singular optics: tailoring optical vortices}
Lately, chiral surface plasmon modes have also been studied in relation to singular optical effects \cite{Babiker,Zhan2009,Halas,Garcia-Vidal}. Indeed, near-field excitations on chiral nano structures have shown to generating orbital angular momentum (OAM) both in the near field \cite{Ohno,PRL2008,KimNanoLett2010,ChoOptX2012} and the far field \cite{Nano2009,Zhan}. We have just recently presented a comprehensive analysis of 
the OAM transfer during plasmonic in-coupling and out-coupling by chiral nanostructures at each 
side of a suspended metallic membrane, stressing in particular the role of a back-side structure in generating vortex
beams as $e^{i\ell\varphi}$ with tunable OAM indices $\ell$. \\
\indent Our device consists of a suspended thin ($h\sim 300$ nm) metallic membrane, fabricated by evaporating 
a metal film over a poly(vinyl formal) resine supported by a transmission electron microscopy 
copper grid. After evaporation, the resine is removed using a focused ion-beam (FIB), leaving a 
freely suspended gold membrane. Plasmonic structures are milled, in either concentric (BE) or spiral
geometry on both sides of the membrane around a unique central cylindrical aperture acting as 
the sole transmissive element of the whole device. The general groove radial path is given in the polar $(\hat{\boldsymbol \rho},\hat{\boldsymbol \varphi})$ basis, 
as ${\boldsymbol \rho}_{n}=(n\lambda _{\rm SP}+m\varphi \lambda _{\rm SP}/2\pi)\hat{\boldsymbol \rho}$, with 
$n$ an integer, $\lambda _{\rm SP}$ the SP wavelength and $m$ a pitch number. Orientation conventions are chosen with respect to the light 
propagation direction, so that a right-handed spiral $R_{m}$ corresponds to $m > 0$ and a left-handed spiral 
$L_{m}$ to $m < 0$. 
As we summarize below, there is an close relation between the spiral pitch $m$ and the topological charge $\ell$ of the vortex generated in the near field.\\
\indent Near-field generation of OAM at the front-side ($z=0^+$) of the structured membrane can be understood by considering that each point ${\boldsymbol \rho}_n$ of the groove illuminated by the incoming 
field is an SP point source, launching an SP wave perpendicularly to the groove. With groove widths 
much smaller than the illumination wavelength, the in-plane component of the generated SP field in 
the vicinity of the center of the structure is ${\bf E}^{\rm SP}({\boldsymbol \rho}_0,z=0^+)\propto {\sf G}\cdot [{\bf E}^{\rm in}({\boldsymbol \rho}_n,z=0^+)\cdot\hat{\bf n}_n]\hat{\bf n}_n$ where ${\sf G}=e^{ik_{\rm SP}|{\boldsymbol \rho}_0-{\boldsymbol \rho}_n|}/(|{\boldsymbol \rho}_0-{\boldsymbol \rho}_n|)^{1/2}$ is the Huygens-Fresnel plasmonic propagator and $\hat{\bf n}_n=\kappa^{-1} ({\rm d}^2{\boldsymbol \rho}_n / {\rm d}s^2)$ the local unit normal vector determined from the curvature $\kappa$ and the arc length $s$ of the groove. The resultant SP field is the integral of elementary point sources over the whole groove structure. As indicated by a full evaluation, we can conveniently limit the integration to radial regions $\rho_n\gg\rho_0$ where the grooves become pratically annular. This leads to $\hat{\bf n}_n \sim -\hat{\boldsymbol \rho}$ and therefore to a simple expression of the integrated SP field ${\bf E}^{\rm SP}={\sf C}_{\rm in}\cdot {\bf E}^{\rm in}$, connected to the incoming field by an in-coupling matrix
\begin{eqnarray}
{\sf C}_{\rm in}(m)\propto e^{im\varphi_0}\int\limits_0^{2\pi}{\rm d}\varphi e^{im\varphi}e^{-ik_{\rm SP}\rho_0\cos \varphi}
\hat{\boldsymbol \rho}\otimes\hat{\boldsymbol \rho},
\label{matIn}
\end{eqnarray} 
the $\otimes$ symbol denoting a dyadic product. \\
\indent Contrasting with the recent studies that have been confined to the near field, our suspended membrane opens the possibility to decouple the singular near field into the far field, with an additional structure on the back-side of the membrane connected to the front-side by the central hole. By symmetry (assuming loss-free unitarity) the out-coupling matrix is simply given as the hermitian conjugate of the in-coupling matrix, i.e. 
${\sf C}_{\rm out}={\sf C}_{\rm in}^{\dagger}$, corresponding to a surface field that propagates away 
from the central hole on the back-side. The in -- out-coupling sequence corresponds to the product ${\sf T}={\sf C}^{\dagger}(m_{\rm out}) \cdot {\sf C}(m_{\rm in})$ which, in the circular polarization basis, writes explicitly as
\begin{eqnarray} 
{\sf T}\propto e^{i(m_{\rm out}-m_{\rm in})\varphi}
\begin{bmatrix}
{\sf t}_{++}     & {\sf t}_{+-}e^{2i\varphi}\\ 
{\sf t}_{-+}e^{-2i\varphi} &    {\sf t}_{--}
\end{bmatrix}
\label{matT}
\end{eqnarray}
with ${\sf t}_{ij}$ radial functions based on products of $m_{\rm in,out}\pm 1$-order Bessel functions of the first kind, as detailed in \cite{Gorodetskisub}. Note that we use here circular basis conventions that allow associating a positive value to an OAM induced on a right $m>0$ right spiral. With respect to this chapter, these conventions are such that $+/-$ is associated with $\hat{{\bf L}}/\hat{{\bf R}}$. \\
\indent This expression (\ref{matT}) reveals two contributions: a polarization dependent geometric phase, within the matrix, that stems from the spin-orbit coupling at the annular groove, and a factorized dynamic phase that arises due to the spiral twist of the structure \cite{PRL2008}. In relation to what has just been described above, the structure of Eq. (\ref{matT}) shows that, it would belong to the general class of chiral structure, as it is not a mere 2D chiral structure, as seen from the fact that ${\sf t}_{++}\neq {\sf t}_{--}$ nor does it belong to the simple optical activity class given that $|{\sf t}_{+-}|\neq |{\sf t}_{-+}|$. Obviously when $m_{\rm in}=m_{\rm out}=0$, the system is achiral and in this case, ${\sf T}$ describes a pure spin--orbit angular momentum transfer, conserving the total angular momentum \cite{PRL2008,Marucci,Brasselet,Manu}.\\
\begin{figure}[h]
\centering\includegraphics[width=10cm]{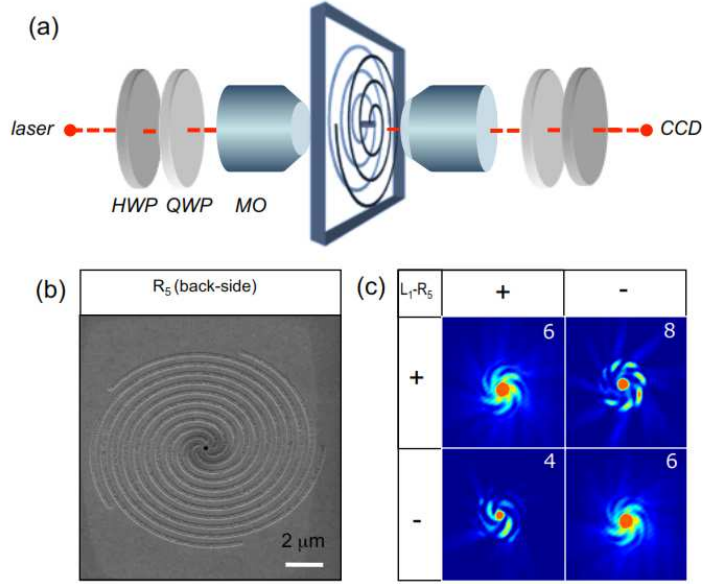} \caption{(a) Experimental setup: the incoming laser beam is circularly polarized using half (HWP) and quarter-wave (QWP) plates and weakle focused by a microscope objective ($5\times$, NA$=0.13$). The transmitted beam is imaged by a second objective ($40\times$, NA=$0.60$) and a lens tube ($f=200$ mm, not shown) on a CCD camera and analyzed in the circular polarization basis by additional HWP and QWP. (b) Scanning electron microscope image of an $R_5$ spiral milled on the back side of a gold membrane, with $\lambda_{\rm SP}=768$ nm. (c) Intensity distribution imaged through a $L_1-R_5$ structure.  Labels ($\pm, \pm$) correspond to the combination of circular polarization preparation and analysis. The numbers correspond to the corresponding OAM indices.
Images adapted from Refs.~\cite{Gorodetskisub}}\label{fig4}
\end{figure}
\indent We have checked experimentally the OAM summation rules that can be drawn from this analysis and that can be gathered in a summation table. A sketch of the experiment is presented in Fig. \ref{fig4}. As also shown in panel (c) of the figure, we have been able to generate optical beams in the far field with OAM indices up to $\ell = 8$, in perfect agreement with the expected OAM summation rules. Note that this $\ell = 8$ value is in strict relation with the chosen structures and is not a limit to our device.\\
\begin{table}[htbp]
	\centering
		\begin{tabular}[t] {|c|c|c|}\hline
		                        & $+$ & $-$ \\ \hline
$+$		&  $m_{\rm out}-m_{\rm in}$            &   $m_{\rm out}-m_{\rm in}+2$         \\ \hline
$-$  	&  $m_{\rm out}-m_{\rm in}-2$          &   $m_{\rm out}-m_{\rm in}$            \\ \hline
	\end{tabular}
	\caption{Far field summation rules for OAM generated through the membrane.}
\label{table}
\end{table}
\begin{figure}[h]
\centering\includegraphics[width=10cm]{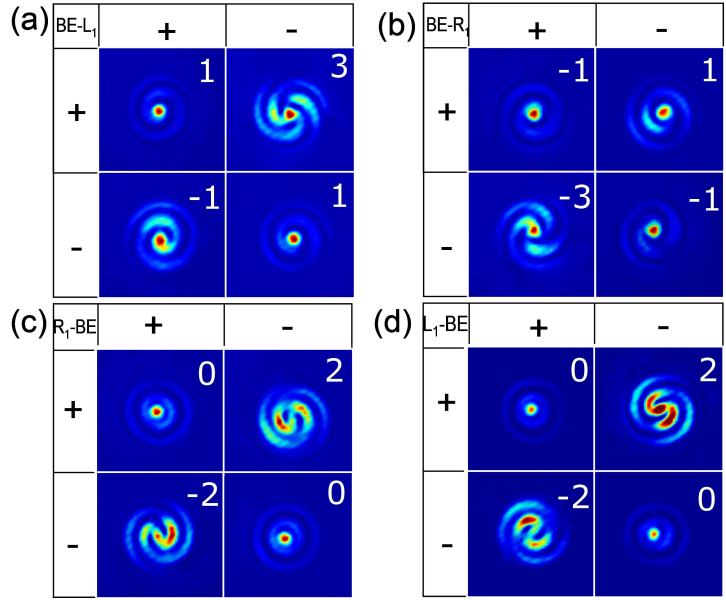} \caption{Intensity distributions of the beam emerging from (a) BE-$L_1$, (b) 
BE-$R_1$, (c) $R_1$-BE and (d) $L_1$-BE structures, respectively. 
The hole diameter used in all the structures was $400$ nm. 
Same labels as in Fig. \ref{fig4}. Images adapted from Refs.~\cite{Gorodetskisub}}\label{fig5}
\end{figure}
\indent Remarkably with suspended membranes, one can actually perform experimentally the path reversal operation described in section \ref{path} and in Fig. \ref{fig2}. By merely flipping the BE-$(L,R)_1$ structures with respect to the optical axis, one obtains $(R,L)_{1}$-BE, with enantiomorphic changes $L\leftrightarrow R$ generated by the planar character of the spirals milled on one side of the membrane. \\
\indent As displayed in Fig.\ref{fig5} when comparing panels (a) and (c) and (b) and (d), the OAM measurements however turn out to be inconsistent with a simple path reversal operation. As we fully explain in \cite{Gorodetskisub}, this discrepency points to the pivotal role of the central aperture in the process of OAM conservation, inducing specific OAM selection rules that must be accounted for in the generation process. This discussion is beyond the scope of this summary, and we refer the reader to our manuscript for further details.

\section{Outlook}
We have tried in this chapter to given an overview on the close relation between the chiral behavior of optical media and reciprocity. This relation has raised recently interesting issues, particularly salient in the context of metamaterial optics. As we discussed here, an algebraic approach can be useful as it allows distinguishing different classes of chiral media with respect to reciprocity. Along these lines, one is naturally led to unveil new types of chiroptical behaviors such as {\it optical planar chirality}, a totally original optical signature that contrasts with standard {\it optical activity}. Doing so, we have also stressed how reciprocity should not be confused with time-reversal invariance. A couple of recent experimental examples have been presented that nicely illustrate these intertwined relations in the realm of nanophotonics, both in terms of polarization dynamics and optical vortices.

\section{Acknowledgments}
The authors are grateful to T. W. Ebbesen for his continuous support and Y. Gorodetski, J.-Y. Laluet, and E. Lombard for their participation in the work described here. This work received the financial support of the ERC (grant 227557). 


\end{document}